# The impact of Industry 4.0 technologies on production and supply chains

Davood Qorbani [1] and Stefan Grösser [1,2]

*(1) Bern University of Applied Sciences, Switzerland; (2) University of St. Gallen, Switzerland*

**Abstract**

This paper sheds light on the current development in major industrialized countries (such as Germany, Japan, and Switzerland): the trend towards highly-integrated and autonomous production systems. The question is how such a transition of a production infrastructure can take place most efficiently. This research uses the system dynamics method to address this complex transition process from a legacy production system to a modern and highly integrated production system (an Industry 4.0 system). The findings mainly relate to the identification of system structures that are relevant for an Industry 4.0 perspective. Our research is the first in its kind which presents a causal model that addresses the transition to Industry 4.0.

## 1. Introduction

What is innovation for some, is disruption for others. One day, we are market disrupters, and the next day we may need to manage the disruption of our rivals. More interestingly, the disruption that we make causes new conditions, triggers actions and reactions from our rivals, our customers, and other stakeholders which ultimately comes back to us in a way that creates a different environment from the one that we started from. In simple words, what we decide and do today is going to meet us tomorrow.

Businesses in general, and more specifically the industrial production are in the midst of a fundamental transformation. This paper addresses a transition to digitalization (i.e. the introduction of advanced information technologies) in energy and production industries considered from a systems (i.e. integrated) perspective. Before introducing the challenge that this paper addresses, we define the two central concepts of *digitalization*, and *Industry 4.0*.

### *Definition: Digitalization*

Although the term digitalization is very broad and an open concept, and is "often used interchangeably with digitization (as well as with digital transformation), […] the term […] has been coined to describe the manifold sociotechnical phenomena and processes of adopting and using [digital] technologies in broader individual, organizational, and societal contexts" (Legner et al., 2017, p. 301). By the term *digitalization* in our study, we mean the wave that has transformed different industries such as the media (Hjarvard & Kammer, 2015; Rothmann & Koch, 2014; Viljakainen, Toivonen, & Seisto, 2016), publications (Park, Lee, & Casalegno, 2010; Picard, 2003) music and recordings (Dolata, 2008), and banking (Sia, Soh, & Weill, 2016), and has overthrown their previously successful dominant business models (cf. Legner et al., 2017). This transition is about to change additional industries, e.g., the energy industry, and the production industry.

### *Definition: Industry 4.0*

Industry 4.0 is a relatively new term, and gained popularity first in Germany - and then in the neighboring countries - after Angela Merkel, German Chancellor, announced the national strategic initiatives in 2011 (*Germany: Industrie 4.0*, 2017). By one definition, Industry 4.0 or the Industrial Internet of Things (Beier, Niehoff, Ziems, & Xue, 2017), is "the convergence of industrial production and information and communication technologies", although "a generally accepted understanding of the term does not exist" (Hermann, Pentek, & Otto, 2016, p. 3928). Industry 4.0 is a collective term (Lesjak, Druml, Matischek, Ruprechter, & Holweg, 2016) and refers to advances in information and communication technologies, the computerization of manufacturing, and a very fast growth of cyber-physical systems (Hermann et al., 2016; Wang, Wan, Zhang, Li, & Zhang, 2016) which have resulted in a synergy that triggered such a fundamental transformation in industrial production systems.

Industry 4.0 is being defined within digitalization and is about production systems, such as the car manufacturing industry or the medical devices production system. Industry 4.0 is physical and product-oriented. In comparison, digitalization is a broader concept and has been applied far beyond the industrial sector. For instance, it extends to the service-sector, the health sector, and the financial sector. Digitalization has also transformed industries outside Industry 4.0, such as the music industry (see Hermann et al. (2016); Lesjak et al. (2016); Stettler (2017)).
.

## 2. Literature Review

Since the concepts of Industry 4.0 and as well as digitalization of industrial productions are new, the literature, and concrete works on the topic are scant. Srai et al. (2016, p. 6917) argued that digitalization "together with infrastructural developments (in terms of the Internet of Things and big data) provide new opportunities." A number of opportunities have been cited for digitalization: it enhances the "capability to manufacture closer to the point of demand, with greater specificity to individual needs" (Srai et al., 2016, p. 6933); it enables decentralized decision making (Marques, Agostinho, Zacharewicz, & Jardim-Goncalves, 2017); and, it potentially modifies "the competitive positioning of companies along the supply chain" (Calabrese, 2011, p. 212). More specifically, Industry 4.0 "can offer a reduction in energy consumption, increase economic benefits, and enable smart production" (Li et al., 2017, p. 23). The smart production comes from the smart factory which is one of the key features of the current major transformation, and is "the vertical integration of various components inside a factory to implement a flexible and reconfigurable manufacturing system" (Wang et al., 2016, p. 158).

Andersen (2006) conducted one of the first studies on the digitalization of production activities and showed how it affect coordination costs. In one study, Wang et al. (2016, p. 158) conducted an agent-based modeling and proposed "an intelligent negotiation mechanism for agents to cooperate with each other". In another study, Beier et al. (2017, p. 227) conducted empirical surveys and compared expectations of managers in manufacturing companies of different sector on the topic of digitalization of industrial activities in Germany and China. According to these scholars, the expectation is "that this transformation will not only impact the ecological dimension (resource efficiency, and renewable energy), but that the technical transformation is likely to be accompanied by social transformations." The social consequences of such a transformation – or in other words, the transition to the digitalized industrial production systems – are multifaceted. For example, robotics has raised concerns regarding the opportunities for employment among those who make a living based on the legacy technologies and procedures (Beier et al., 2017). It is foreseeable that "new fields of activity may arise" consequently (Kubinger & Sommer, 2016, p. 330), but that calls for lifelong education and training of the existing workforce.

Moreover, "Industry 4.0 processes including digitalization of industry are game-changing developments in economic evolution" (Schlogl, 2017, p. 533). The strong wave of the change has left obsolete business models, and forced businesses that have been reluctant or resistant to such change to close. The business digitalization, in general, puts the business models of pre-digitalization at risks, and has the potentials to threat the position of market leaders. It is foreseeable that the digitalization will write a similar destiny for the businesses of the industrial production sector. For instance, prosumers constitute a new generation of customers who not only consume products and makes use of services, but also have become producers and are involved in the process of designing and manufacturing products (Stahel, 2016). Thus, they constitute a major challenge to the conventional top-down business model of industrial production systems. For example, in the electricity sector – an industrial production system – the digitalization has been triggered already. A decentralized business model of electricity production – in which household consumers can produce electricity to not only cover their electricity needs, but also sell the excess to the power grid – challenges the conventional and familiar business models of electricity supply by utility companies (Curtius, Künzel, & Loock, 2012; Giordano & Fulli, 2012). Industry 4.0, has enabled such a bilateral flow of electricity, and interactions between consumers and utility companies.

## 3. The Addressed Challenge

The competition in industries is fierce and evermore intensifying. The companies and businesses are under an immense pressure to respond to these increasing changes in the demand from customers, and the speed by which those demands need to be answered. Companies in each industry crave for solutions that support their daily operations and allow them to remain competitive.

The transition to Industry 4.0 is costly, because it requires a fundamental change in the business operations of the companies that embark on such a transition. Furthermore, a renewal of knowledge that exist within a company, as well as an increase in the qualifications required of workers and employees are unavoidable (Beier et al., 2017). Last but not least, the transition entails risks. Large investments in facilities, fundamental changes in the workforce mix, the introduction of new business models, acceptance and adaptation by customers (consumers and/or prosumers) and by the society at large, and a sociotechnical aspect, implies the risks that may threaten the profitability of such businesses (Beier et al., 2017).

This paper is a part of an ongoing research to investigate the performance improvement that introducing Industry 4.0 and the synergy among its various components promises to a production and supply chain of a Swiss medium-sized industrial production system. At this milestone and in this paper, we consider the core functions of an industrial production system which is supposed to embark on the transition to Industry 4.0. We seek to answer the following research question: What are the underlying dynamics that govern the transition from the legacy production systems to Industry 4.0 which creates the performance improvement in a production system (production line)?

## 4. Analysis Method

Our research strategy on this topic would be performing a case study "in order to obtain a clear picture of [the] problem … from various angles and perspectives using multiple methods of data collection" (Sekaran & Bougie, 2016, p. 98). Our case will be a Swiss medium-sized production-oriented company.

In this study, we want to detail how digitalization – and more specifically, Industry 4.0 – influences a production-oriented company. We select a medium-sized company for two reasons. First, this category of companies is in an uncovered territory; almost all big companies already made or are in the process to transition to Industry 4.0 to reap the benefits. The management of medium-sized companies lacks knowledge about Industry 4.0, what changes it involves, and what costs and benefits may result. Most of the small-sized companies also do not care about the industry 4.0. This is primarily because the size of the impact may not be considerable/relative to the required investment. Engaging in an Industry 4.0 transition is quite expensive and if an industrial production system deals with small-batches of production, it may not be profitable for the firm to undergo such a transition. Furthermore, the transition needs a large, costly repository of knowledge. When the production volume is too low and the required knowledge is not available, then a small company will not profit from the Industry 4.0.

The second reason for focusing on the medium-sized companies is that this category accounts for slightly more than one-fifth of the employment in Switzerland, and, combined with large enterprises, it accounts for more than half of the employment in the country (Federal Statistical Office, 2017). An Industry 4.0 transition in medium-sized companies, consequently will cause the Swiss business environment to be re-drawn.

### *Data Sources*

The initial sources of data for this case study exists on the Internet, such as reports, articles, and cases on production-oriented companies, as well as statistical data from the Swiss Federal and public databases.

### *Method*

Delays, originating from underlying accumulation processes, positive and negative feedback processes, nonlinearities, and uncertainty governs our case on "the impact of Industry 4.0 technologies on production and supply chains". Below are some case-specific examples of such components:

Building the required knowledge and training staffs with different technologies in Industry 4.0 is a gradual one and takes place over time. So is the transition of non-capable to Industry 4.0 capable machinery and the transition of non-capable to Industry 4.0 capable human resources.

Underlying the dynamics of this study, both reinforcing (positive) and attenuating (negative) feedback loops can be identified. For example, larger sales positively impact the revenue streams in the production and supply chains. Larger revenue streams may be used for capital investments on new technology and modern machineries. Such machineries induce product quality which in turn creates value. The value may positively impact the revenue stream. This closes one positive feedback loops. Such positive (reinforcing) feedback processes may interact with the negative (balancing) one: Operationalizing new machines and the production may be hampered by the low literacy of employees. Lower operationalized machineries may drain the value from the company. It is from the interaction between such underlying reinforcing and balancing feedback processes that the dynamics emerges. In this example, we see the need for coordination between policies in areas of personnel, and capital acquisition.

Furthermore, we see the presence of non-linearity in the in the above example. Lack of trained employees shuts down the production line of a modern production and supply line, despite the level of investment in modern machineries. In addition, there is uncertainty on the level and the speed of training that employees experience, or uncertainty on the adequate number of machineries for the trained staffs. Digitalized production machinery may not produce efficiently or even effectively without qualified personnel. Industry 4.0 qualified personnel may not be as productive as expected without digitalized capital. The implication in this example is the need for coordination between policies in areas of personnel, and capital acquisition.

The method that we will use to achieve the objectives of this study, and answer the research question of the topic would be system dynamics, - modeling, simulation and analysis. It addresses systems exhibiting dynamics that originate from the underlying structural properties outlined above; accumulation processes causing delays, feedback of information and material, non-linearity causing synergy between systems components and uncertainty (cf. Meadows & Wright, 2009; Morecroft, 2015; Sterman, 2000). This method enables us to synthesize our empirical evidence into an integrating model, a consistent and coherent repository of knowledge that allows for simulation and analysis of the intimate relationship between model structure and dynamics. The method has been widely applied in the energy sector (Radzicki & Taylor, 1997; Spataru, 2017) as well as environmental and sustainability sciences (Ford, 2009).

## 5. Basic Model

Figure 1 presents the core message of this paper. Because of the advantages that is being accompanied with an Industry 4.0 production system (as mentioned in the Literature Review), this is the case that cost per unit of production by non-capable Industry 4.0 machines is higher than that of the Industry 4.0 capable machines. Thus, transition to Industry 4.0 reduces the production costs. However, this is not the whole story. As discussed earlier, the transition is costly and disrupts the production process of a typical company. A large size cost of the transition may bankrupt the firm. Besides, more dynamics are in play in such a transition and different policy areas should be organized in harmony to synergize for a successful transition.

## 6. Model of the Transition to Industry 4.0

Figure 2 Presents a model of the transition to Industry 4.0 which incorporates larger dynamics compared to the basic model which we presented in the previous section (some variables are removed from this figure for the demonstration purpose).

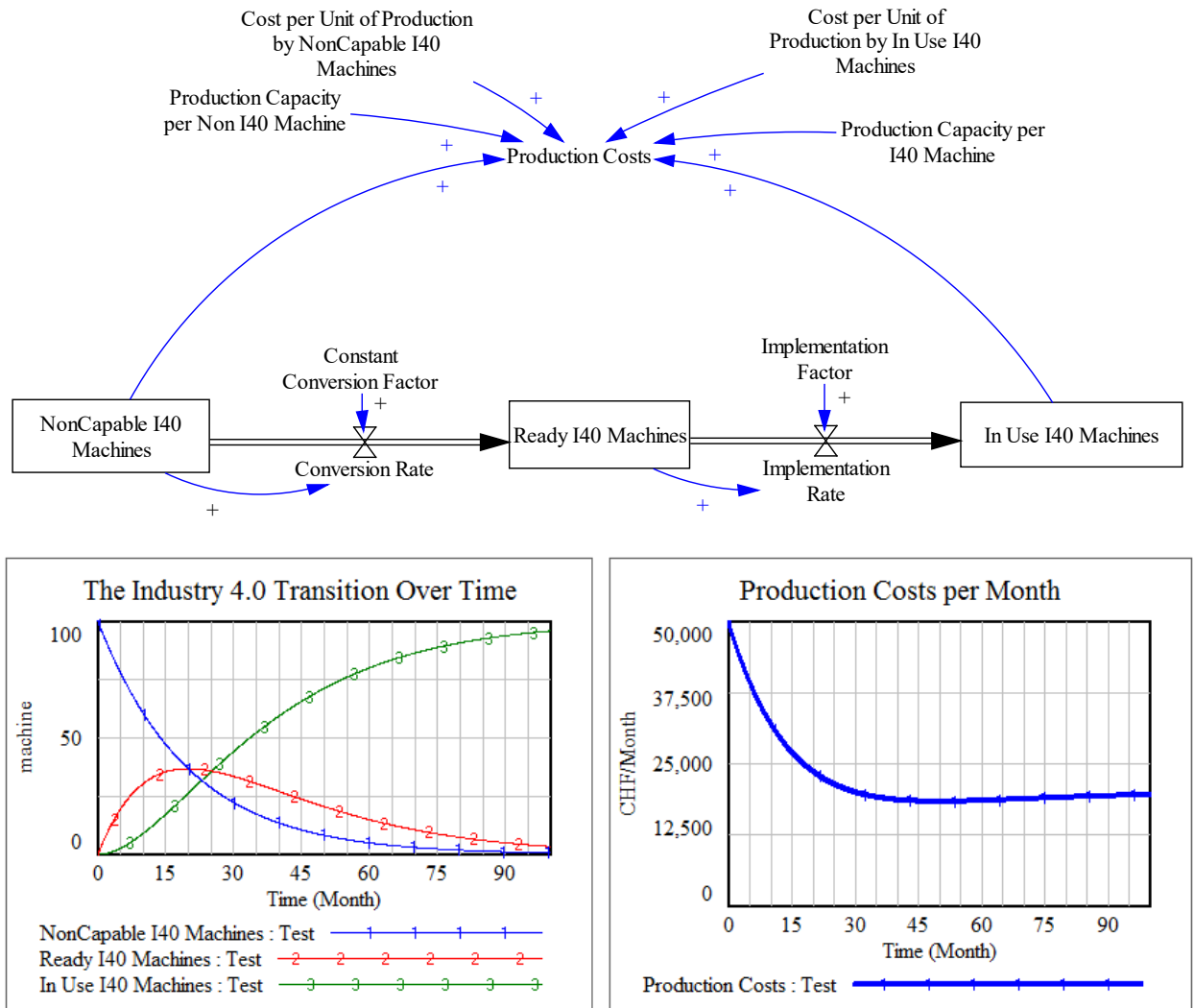

*Figure 1: Production Costs in Transition to Industry 4.0*

Here we describe four important feedback loops in our larger model. The first loop, a balancing one, is about hiring and dismissal of staffs. For the simplicity in the model, we overlooked the fact the current employees or staff could be trained to work with high-tech Industry 4.0 machines. Dismissal of those types of staff puts financial burden on the company, because financial compensations is required according to the law. At the other hand, employing trained IT staff or technicians who usually ask for higher payments costs money for the company. Both dismissal and hiring negatively affect financial liquidity of the company which ultimately hampers the acquisition of new Industry 4.0 machineries.

The other balancing loop is the acquisition of high-tech machineries. Although these types of machineries are not labor-intensive during production, however acquisition of them is capital-intensive and drains the cash from the company and slows down the acquisition. The next balancing loop is a side-effect of the dynamics of acquisition. The ready, but not in-use Industry 4.0 machines needs to be kept and take care of which usually incurs cost of storage which in turn drain liquidity of the company.

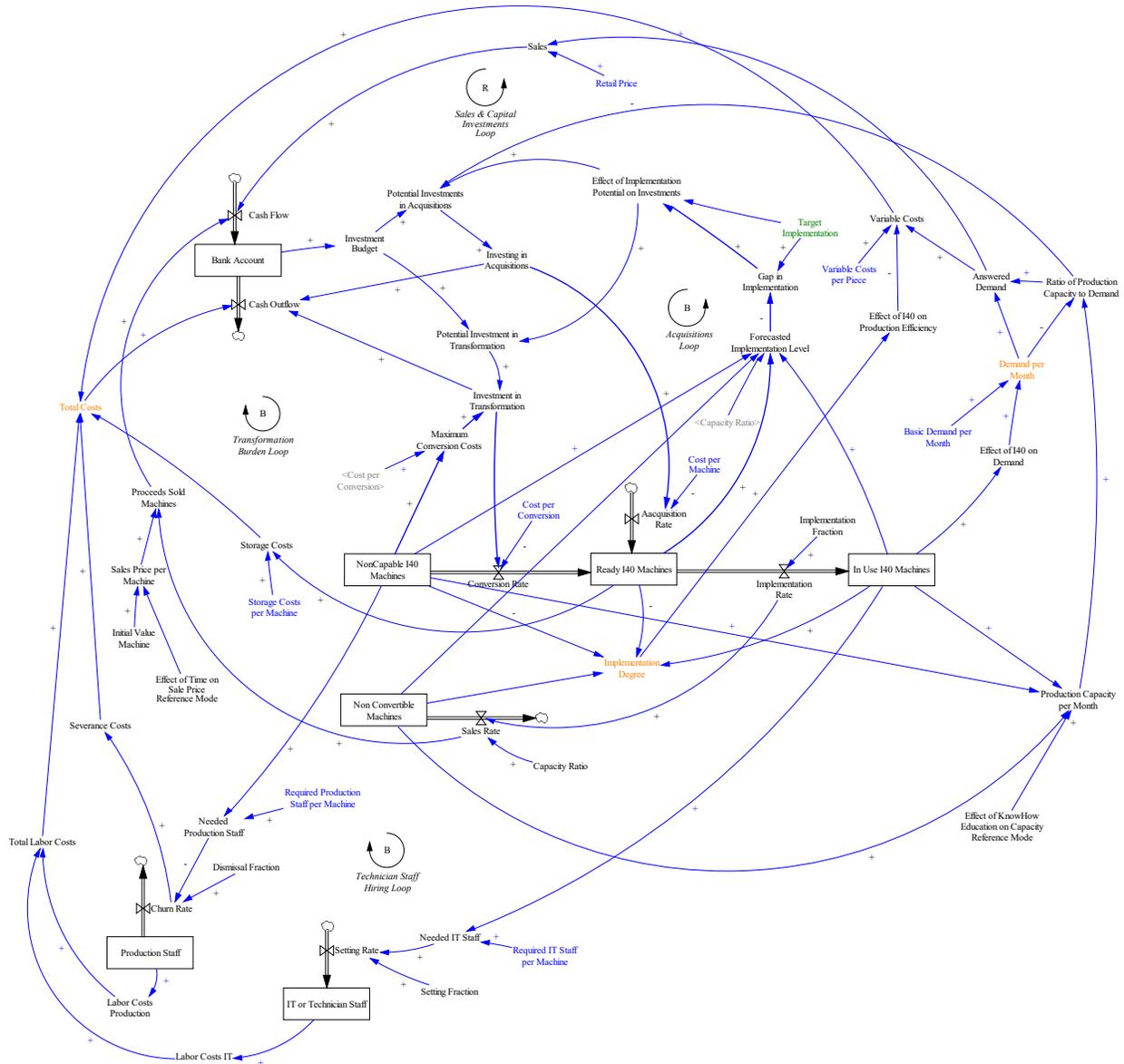

*Figure 2: A Model of the Transition to Industry 4.0*

Finally, there is a reinforcing loop of sales and capital investment. Larger sales positively impact the revenue streams in the production and supply chains. Larger revenue streams may be used for capital investments on new technology and modern machineries.

The above-mentioned feedback loops interact and synergize in creating behaviors that is being observed in different policy areas of a company that embarks on the transition to Industry 4.0. Figure 3, for example, shows that how the combination of the staff will change over the transition period in such a company.

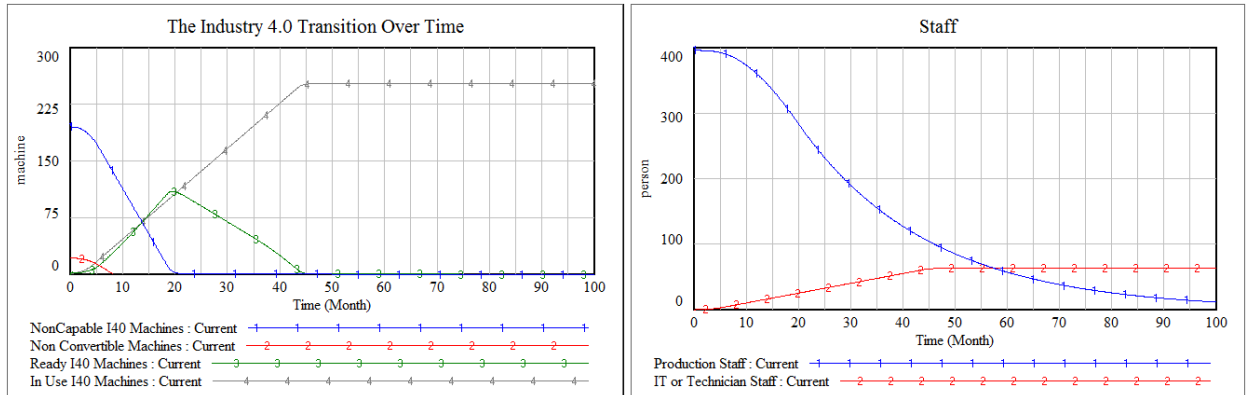

*Figure 3: Change in the Combination of the Staff over the Transition*

## 7. Concluding Remarks

This paper aims to explore the underlying structure of the behavior that governs the transition from the legacy production systems to highly integrated production system of Industry 4.0, and considers the probable risks which comes with such a transition.

The contribution of this study is two-fold. First, it contributes with new literature on digitalization of industrial production and, more specifically, on the impact of Industry 4.0 on medium-sized companies. Second, it provides the company of the case study with an integrated perspective on different inter-related key performance indicators (KPIs) of its business, such as the level of cash, and the cash flow; human resources employed and staffed on the legacy and new machines. This is important because the dynamic effects of the policies that govern decision making in various parts of an enterprise, synthesize due to non-linearities, in creating the kind of behavior we observe. Thus, policy design and the resulting decision making will need to be coordinated.

Our next step is to extend the current model to incorporate more dynamics of the mentioned transition in industrial production systems. Digitalization of industrial production systems, or in other words Industry 4.0, seems to be promising in eliminating bottlenecks that hinders the production in legacy production systems. Such bottlenecks in production lines come from various types of constraints such as financial problems, or time pressures (Pegels & Watrous, 2005; Rahman, 1998). Prioritizing where needs to be loosen up and getting leverage from eliminating the mentioned bottlenecks, or in other words de-bottlenecking, and the consequent effect on the different aspect of a product is important in production lines.